\documentclass[11pt,twoside]{article}
\usepackage{./asp2014}

\aspSuppressVolSlug
\resetcounters

\begin{document}

\title{Cold gas in High-z Galaxies: CO as redshift beacon}
\author{R.~Decarli,$^{1,2}$ C.~Carilli$^{3,4}$, C.~Casey$^5$, B.~Emonts$^6$, J.A.~Hodge$^7$, K.~Kohno$^8$, D.~Narayanan$^9$, D.~Riechers$^{10}$, M.T.~Sargent$^{11}$, F.~Walter$^{2,3}$
\affil{$^1$INAF -- Osservatorio di Astrofisica e Scienza dello Spazio, via Gobetti 93/3, 40129 Bologna, Italy; \email{roberto.decarli@inaf.it}}
\affil{$^2$Max Planck Institut f\"{u}r Astronomie, K\"{o}nigstuhl 17, 69117 Heidelberg, Germany}
\affil{$^3$National Radio Astronomy Observatory, Pete V.~Domenici Array Science Center, P.O.~Box O, Socorro, NM 87801, USA}
\affil{$^4$Cavendish Laboratory, University of Cambridge, 19 J.J.~Thomson Avenue, Cambridge CB3 0HE, UK}
\affil{$^5$Department of Astronomy, The University of Texas at Austin, 2515 Speedway Blvd Stop C1400, Austin, TX 78712}
\affil{$^6$National Radio Astronomy Observatory, 520 Edgemont Road, Charlottesville, VA 22903, USA}
\affil{$^7$Leiden Observatory, Niels Bohrweg 2, 2333 CA Leiden, The Netherlands}
\affil{$^8$Institute of Astronomy, School of Science, The University of Tokyo, 2-21-1 Osawa, Mitaka, Tokyo 181-0015, Japan}
\affil{$^9$Department of Astronomy, University of Florida, 211 Bryant Space Science Center, Gainesville, FL 32611, USA}
\affil{$^{10}$Cornell University, 220 Space Sciences Building, Ithaca, NY 14853, USA}
\affil{$^{11}$Astronomy Centre, Department of Physics and Astronomy, University of Sussex, Brighton, BN1 9QH, UK}
}

\paperauthor{R.~Decarli}{roberto.decarli@inaf.it}{0000-0002-2662-8803}{INAF}{Osservatorio di Astrofisica e Scienza dello Spazio}{Bologna}{BO}{40129}{Italy}
\paperauthor{C.~Carilli}{ccarilli@nrao.edu}{}{NRAO}{Pete V.~Dominici Array Science Center}{Socorro}{NM}{87801}{USA}
\paperauthor{C.~Casey}{cmcasey@astro.as.utexas.edu}{}{University of Texas at Austin}{Department of Astronomy}{Austin}{TX}{78712}{USA}
\paperauthor{B.~Emonts}{bjornemonts@gmail.com}{}{National Radio Astronomy Observatory}{50 Edgemont Road}{Charlottesville}{VA}{22903}{USA}
\paperauthor{J.A.~Hodge}{hodge@strw.leidenuniv.nl}{}{Leiden Observatory}{}{Leiden}{CA}{2333}{The Netherlands}
\paperauthor{K.~Kohno}{kkohno@ioa.s.u-tokyo.ac.jp}{}{School of Science, University of Tokyo}{Institute of Astronomy}{Tokyo}{Mitaka}{181-0015}{Japan}
\paperauthor{D.~Narayanan}{desika.narayanan@gmail.com}{}{University of FLorida}{Department of Astronomy}{Gainesville}{FL}{32611}{USA}
\paperauthor{D.~Riechers}{riechers@astro.cornell.edu}{}{Cornell University}{}{Ithaca}{NY}{14853}{USA}
\paperauthor{M.T.~Sargent}{Mark.Sargent@sussex.ac.uk}{}{University of Sussex}{Astronomy Centre, Department of Physics and Astronomy}{Brighton}{}{BN1 9QH}{UK}
\paperauthor{F.~Walter}{walter@mpia.de}{}{MPIA}{Galaxies and Cosmology}{Heidelberg}{}{69117}{Germany}


\section{Science Goals}
The goal of this science case is to address the use of a ngVLA as a CO redshift machine for dust-obscured high-redshift galaxies which lack of clear counterparts at other wavelengths. Thanks to its unprecedentedly large simultaneous bandwidth and sensitivity, the ngVLA will be able to detect low--J CO transitions at virtually any $z>1$. In particular, at $z>4.76$ two CO transitions will be covered in a single frequency setting, thus ensuring unambiguous line identification. The ngVLA capabilities fill in a redshift range where other approaches (e.g., photometric redshifts, search for optical/radio counterparts, etc) typically fail due to the combination of intrinsically faint emission and increasing luminosity distance. This will allow us to explore the formation of massive galaxies in the early cosmic times.


\section{Scientific rationale}

Since the identification of the first sub-mm galaxies and dusty star-forming galaxies \citep[DSFGs; e.g.,][]{hughes98,ivison98,ivison02,smail02,chapman05}, it has been clear that obtaining precise redshift measurements for these sources would represent a major challenge, especially at the highest redshifts \citep[e.g.,][]{riechers13,marrone18}. A combination of intrinsic faintness, high redshift, and dust reddening limits the detectability of rest-frame optical/UV spectral features that could be used to pin down the redshift to $\sim$50\% of the DSFGs studied so far \citep[e.g.,][]{danielson17}. This is particularly severe for those galaxies for which the estimated unobscured star formation is only a tiny fraction of the total star--formation rate. The peak wavelength of the dust spectral energy distribution can be used as a tentative redshift proxy \citep[see, e.g.,][]{riechers13,dacunha15}, although it is highly degenerate with the dust temperature. The search for radio continuum counterparts biases the selection against high redshifts (as the radio continuum does not benefit of the negative $k$-correction that is in place at sub-mm wavelengths). As a result, the intrinsic redshift distribution of these sources is still under debate \citep[see][for a review]{casey14}. The question of whether the unconfirmed DSFGs sit at similar redshifts or higher redshifts than the others is yet to be clarified, with the incompleteness being a strong function of redshift. The discovery of bright dusty galaxies like HFLS3 \citep{riechers13} and SPT0311--58 \citep{marrone18}, and the four [C{\sc ii}]--bright galaxies in the field of four $z>6$ quasars \citep{decarli17} hint at the existence of such a population at very high redshifts, but so far,  confirming such sources (especially at lower luminosities or in the absence of gravitational lensing) in a systematic way has not been efficient. 

A promising way forward consists in searching for multiple CO transitions. Since DSFGs are very luminous at IR wavelengths, and the CO luminosity correlates with $L_{\rm IR}$ \citep[see, e.g.,][]{carilli13}, CO transitions are expected to be bright in DSFGs. CO transitions have rest frequencies of $\sim$115.2 J$_{\rm up}$ GHz. The detection of two transitions might leave some degeneracy in the redshift interpretation [e.g., if one line is observed at 2$\times$ the frequency of the other, they could be CO(1-0) and CO(2-1), or CO(2-1) and CO(4-3), or CO(3-2) and CO(6-5), etc]. However, by combining the line detection with ancillary information (e.g., broad-band optical/NIR photometry) one could significantly restrict the number of possible identifications. Furthermore, the CO spectral energy distribution is typically fairly regular, so that if one observe, e.g., CO(2-1) and CO(4-3), the flux of the CO(3-2) can be predicted pretty accurately. The presence or lack of intermediate transitions can then further trim the parameter space of allowed interpretations. 
\begin{figure}
\centering
\includegraphics[width=0.99\textwidth]{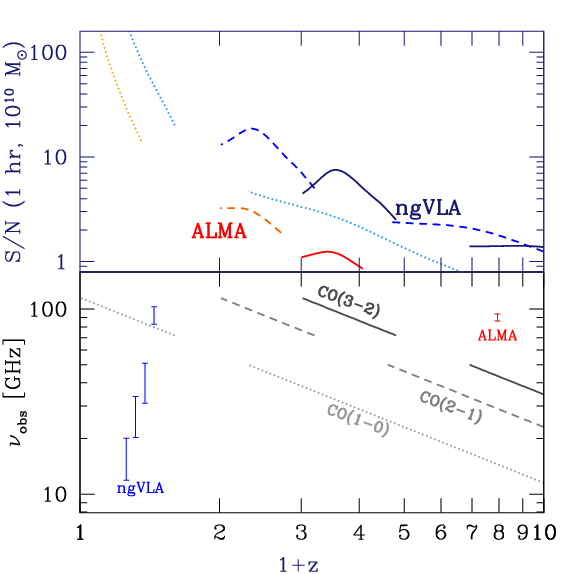}
\caption{The power of ngVLA as a redshift beacon, compared to ALMA. {\em Top ---} The expected S/N of CO(1-0) (dotted), CO(2-1) (dashed), and CO(3-2) (solid lines) emission for a galaxy with molecular mass $M_{\rm H2}=10^{10}$\,M$_\odot$, as a function of redshift. We assume 1\,hr of integration, a line width of 300\,km\,s$^{-1}$, a CO--to--H$_2$ conversion of $\alpha_{\rm CO}$=3.6 M$_\odot$ (K\,km\,s$^{-1}$\,pc$^2$)$^{-1}$ \citep{daddi10}, and a CO excitation typical of main sequence galaxies \citep{daddi15}. The ngVLA (blue lines) will provide 5--10$\times$ higher S/N for the same integration time than ALMA (red lines) in the overlapping frequency ranges, and will expand the redshift coverage of low--J CO transitions to much higher redshifts, virtually covering every redshift $z>1$. {\em Bottom ---} The observed frequency of the same CO transitions as a function of redshift, compared with the simultaneous bandwidth of ngVLA and ALMA (shown as vertical bars). The large bandwidth of ngVLA will enable to sample wide CO redshift ranges in a single frequency setting.}
\label{fig_wg3_redshift}
\end{figure}

The main drawback of CO-based redshifts is that large frequency ranges need to be sampled in order to find one or more lines. At any redshift, CO transitions are spaced by several tens of GHz in the observed frame --- thus representing an observational challenge, as multiple frequency setting need to be combined in order to effectively sample a large bandwidth (see Fig.~\ref{fig_wg3_redshift}). \citet{weiss13} used  existing capability of ALMA to scan a large fraction of the 3mm band, and searched for high-J CO emission in 26 DSFGs. Each source was observed in 5 different frequency settings, for a total of $\sim$10 min on source. However, the high detection rate ($\sim$90\% of sources have been detected in at least 1 CO transition) of such short observations was only possible thanks to the extreme brightness of the strongly-magnified sources in their sample. A similar effort with PdBI targeting the non-lensed galaxy HDF850.1 resulted in the first redshift determination of this archetypal source ($z$=5.183); this however required $\sim$100 hr of integration \citep[e.g.,][]{walter12}. Even with full ALMA, redshift scans are going to be expensive in ``normal'' DSFGs that are not lensed. Additionally, by focusing on wavelengths equal to or larger than 3mm, we can only sample intermediate to high J CO transitions, thus the method would be applicable only to highly excited sources.

The ngVLA is going to revolutionize this line of search, as 1) the large simultaneous bandwidth coverage will maximize the probability of detecting the CO(1-0) line at $z>2$ from any source with only a few frequency settings, even if only very coarse constraints on the redshift are available; 2) by focusing on the lowest-J transitions, it will be insensitive to excitation bias, i.e., the method will be applicable to any CO-emitting gaseous reservoir, not only to the most excited; 3) at $z>4.76$, it will provide coverage of both the CO(1-0) and CO(2-1) lines. This will provide unambiguous redshift identification for all the DSFGs at the highest redshift, where the optical/NIR spectroscopy and the radio continuum follow-up are extremely incomplete. 

This is quantitatively demonstrated in Fig.~\ref{fig_wg3_redshift}, where we show the expected S/N of CO(1-0), CO(2-1), and CO(3-2) in a galaxy with molecular gas mass $M_{\rm H2}$=$10^{10}$\,M$_\odot$, line width $\Delta v$=300\,km\,s$^{-1}$, a CO--to--H$_2$ conversion factor $\alpha_{\rm CO}$=$3.6$ M$_\odot$\,(K\,km\,s$^{-1}$\,pc$^2$)$^{-1}$ \citep{daddi10}, and a CO excitation typical for main sequence galaxies \citep{daddi15}. In only a few hours of integration, the ngVLA will be able to securely detect CO(1-0) in two wide redshift windows ($z<0.5$ and $2.2<z<10$), as well as CO(2-1) and/or CO(3-2) at virtually any redshift $z>1$. For a comparison, ALMA will only target intermediate to high J transitions at $z>3$, and their detection will require several hours per frequency settings, and multiple frequency settings to compensate for the smaller bandwidth. The ngVLA will therefore represent the main facility for CO redshift searches in DSFGs over most of the cosmic history in the future decades. For a comparison, the Square Kilometer Array will only be competitive with ngVLA in this line of search if it will have a dedicated high--frequency (5--50\,GHz) band, which is still under debate: The intermediate frequency band (0.3--10\,GHz) becomes sensitive to CO only at $z>10.5$, where we do not expect a significant population of dusty galaxies to be already in place, and other redshift tracers than CO (e.g., fine structure lines at rest--frame far--infrared wavelengths, which are shifted in the ALMA 1--3\,mm bands at these redshifts) become more effective redshift tracers.

\subsection{Measurements Required}

We need to pin down the redshift of dusty galaxies with unknown or poorly constrained redshift via the observation of the CO(1-0) [and possibly CO(2-1)] line.

\subsection{Uniqueness to ngVLA Capabilities (e.g., frequency coverage, resolution, etc.)}

Thanks to the large simultaneous band coverage, the ngVLA will be a unparalleled facility to blindly search for lines associated with dusty, high-redshift galaxies. For any galaxy at $z>2$, the coverage of one/two lines is ensured irrespective of the redshift. By targeting the lowest J transitions, this line search circumvents any potential excitation bias (that would inevitably affect a similar endeavor with, e.g., ALMA). Finally, thanks to the improved sensitivity with respect to the VLA, this approach will become applicable even to relatively CO-faint sources, whereas to date it has only been used for the very brightest sources.

\subsection{Longevity/Durability: with respect to existing and planned ($>$2025) facilities}

The advent of JWST and the 30m class telescopes will significantly push the sensitivity of optical/NIR/MIR spectroscopic investigations, thus mitigating the spectroscopic incompleteness that currently prevents us from characterizing DFSGs, especially at high redshift. However, even these facilities will have a hard time in measuring redshifts of the reddest sources at the dawn of galaxy formation. For these galaxies, the ngVLA will play a unique role as a redshift machine.

\acknowledgements ...  



\end{document}